\documentclass[iop]{emulateapj-rtx4}

\usepackage{graphicx}  
\usepackage{dcolumn}   
\usepackage{bm}        
\usepackage{amsfonts,amsmath,amssymb,mathrsfs}
\usepackage{color}

\usepackage{natbib,times}
\usepackage{hyperref}
\hypersetup{
  colorlinks=true,        
  linkcolor=blue,         
  citecolor=cyan,         
}

\newcommand{\rmn}[1]{{\mathrm{#1}}}                  
\newcommand{\bmath}[1]{{\boldsymbol{#1}}}

\shorttitle{An upper bound from helioseismology on the SBGW} 
\shortauthors{Siegel \& Roth}

\begin{document}

\title{An upper bound from helioseismology on the stochastic\\
  background of gravitational waves}

\author{Daniel M. Siegel\altaffilmark{1} and Markus Roth\altaffilmark{2}}

\altaffiltext{1}{Max Planck Institute for Gravitational Physics
  (Albert Einstein Institute), Am M\"uhlenberg 1, D-14476 Potsdam-Golm, Germany}
\altaffiltext{2}{Kiepenheuer Institut f\"ur Sonnenphysik, Sch\"oneckstr. 6, D-79104
 Freiburg, Germany}

\begin{abstract}
The universe is expected to be permeated by a stochastic background
of gravitational radiation of astrophysical and cosmological
origin. This background is capable of exciting
oscillations in solar-like stars. Here we show that solar-like
oscillators can be employed as giant hydrodynamical detectors for such
a background in the $\mu\mathrm{Hz}$ to $\mathrm{mHz}$ frequency
range, which has remained essentially unexplored until today. We
demonstrate this approach by using
high-precision radial velocity data for the Sun to constrain the
normalized energy density of the stochastic gravitational-wave background around
0.11\,mHz. These results open up the
possibility for asteroseismic missions like \emph{CoRoT} and \emph{Kepler} to probe fundamental physics.
\end{abstract}
		

\section{Introduction}

A stochastic background of gravitational waves (SBGW) is expected to result
from the incoherent superposition of gravitational radiation from a
large number of sources \citep{Maggiore2000,Sathyaprakash2009}. It can
be subdivided into a cosmological
background of primordial gravitational waves (GWs) generated by processes of
the very early universe, and into an astrophysical background produced
by a large number of unresolved astrophysical sources that consist of an
accelerated mass distribution with a quadrupole moment. Many potential
processes of the very early universe have been proposed as generators
of cosmological SBGWs, examples of which are standard inflationary
models, pre-Big-Bang models, and cosmic strings
\citep{Maggiore2000}. Candidate sources to produce the
astrophysical component are, for instance, compact binary star
systems, core collapse supernovae, and rotating neutron stars
\citep{Schneider2010,Regimbau2011,Amaro-Seoane2013}. Since the
universe has been essentially transparent to gravitational radiation
from the very beginning, GWs are ideal carriers of
information on the physical processes that generated them and thus on
the state of the universe at the time they were produced. This
background encodes information on the very early epoch as well as on
the more recent history of the universe that is not accessible to conventional
astronomical observations based on electromagnetic waves
\citep{Maggiore2000,Sathyaprakash2009,Schneider2010,Hils1990}. Extracting
this astrophysical and cosmological information from the
SBGW and thus disentangling the various contributions requires
GW experiments in a variety of frequency bands.

Tight upper bounds on the SBGW have been placed at high and low
frequencies by Earth-based interferometric detectors such as LIGO (around
$100\,\mathrm{Hz}$; \citealt{Abbott2009}), pulsar timing observations
based on the residuals of pulse arrival times (between
$10^{-9}\!-\!3\times 10^{-8}\,\mathrm{Hz}$; \citealt{Jenet2006}), and cosmic
microwave background (CMB) observations
at large angular scales which indicate an upper limit on the
cosmological component of the SBGW at larger wavelengths than the
horizon size at the time the CMB was produced
($3\times10^{-18}\!-\!10^{-16}\,\mathrm{Hz}$; \citealt{Maggiore2000}). However, the
intermediate frequency regime between $10^{-8}\!-\!10\,\mathrm{Hz}$
has proven to be particularly challenging to be probed by
GW experiments and has thus remained essentially
unexplored until today. Apart from the frequency-independent indirect
limits from CMB and Big-Bang nucleosynthesis (BBN) data \citep{Smith2006,Cyburt2005},
which only place tight bounds on the cosmological
component of the SBGW, there are limits from Doppler tracking of the
Cassini spacecraft ($10^{-6}\!-\!10^{-3}\,\mathrm{Hz}$;
\citealt{Armstrong2003}), from precision orbital monitoring of the
Hulse-Taylor binary pulsar (at $10^{-4}\,\mathrm{Hz}$;
\citealt{Hui2013}), from seismic data of the Earth
($0.05\!-\!1\,\mathrm{Hz}$; \citealt{Coughlin2014}), and from a pair of
torsion-bar antennas (TOBAs; $0.035-0.830\,\mathrm{Hz}$;
\citealt{Shoda2014}). In the $10^{-4}\!-\!10\,\mathrm{Hz}$ regime, in
particular, the astrophysical component of
the SBGW could well outshine the cosmological component, revealing
rich astrophysical information on,
for example, the physics of compact objects and star formation history 
\citep{Regimbau2011,Schneider2010,Hils1990}. Proposed space missions to
explore the latter frequency range are the New Gravitational wave
Observatory NGO (a.k.a. eLISA; \citealt{Amaro-Seoane2013}) and the
DECi-hertz Interferometer
Gravitational wave Observatory DECIGO \citep{Kawamura2011}.

Recent theoretical work on the excitation of global stellar
oscillations by GWs \citep{Siegel2010,Siegel2011} together with
helio- and asteroseismic data allows us to employ the Sun and other
solar-like stars as astronomical test masses for the detection of
GWs. Helioseismic Sun-as-a-star data for internal gravity and pressure
modes of the Sun is available from, e.g., the Global Oscillation at
Low Frequency (GOLF) instrument aboard the \emph{Solar and Heliospheric
Observatory} (\emph{SOHO}) spacecraft \citep{Gabriel1995}, while asteroseismic data
from missions like \emph{CoRoT} \citep{Baglin2006} and \textit{Kepler}
\citep{ChristensenDalsgaard2009} offer the possibility to employ
other solar-like stars.

Here, we propose a method to directly detect or constrain both the
astrophysical and cosmological component of the SBGW at
$\mathrm{mHz}$ and $\mu\mathrm{Hz}$ frequencies by using asteroseismic
observations (Section~\ref{sec:method}). While this method is general and can be
applied to any star, we demonstrate the feasibility of this approach
by deducing a direct upper bound around $0.11\,\mathrm{mHz}$ using our
Sun as a hydrodynamic detector (Section~\ref{sec:results}). Section~\ref{sec:discussion}
is devoted to discussion and conclusions.

\section{Methodology}
\label{sec:method}

Our method is based on the theoretical result that stellar
oscillations can be excited by an SBGW
\citep{Siegel2010,Siegel2011}. In the case of
the Sun, the resulting amplitudes can be close to or comparable with values
expected from excitation by near-surface convection
\citep{Siegel2011}, which is considered to be the main driving
mechanism for oscillations in the Sun and
solar-like stars (e.g.,
\citealt{Goldreich1977,Balmforth1992,Goldreich1994,Samadi2001,Belkacem2008}). The
physical mechanism underlying the excitation by an SBGW is the fact
that GWs manifest themselves in oscillating tidal forces: they
periodically stretch and compress the spatial dimensions orthogonal to
the direction of propagation in a quadrupolar pattern,\footnote{This is in general a
  superposition of the two polarization states of a GW, the cross and
  plus polarization.} thus imposing stresses on the
matter they pass through. Global stellar oscillations can be
classified into acoustic pressure modes (\textit{p} modes), with pressure
gradients as the restoring force, and interior gravity modes (\textit{g} modes),
with buoyancy as the restoring force \citep{Aerts2010,Unno1989}. Both \textit{p}
and \textit{g} modes can be excited by GWs, although only quadrupolar
eigenmodes can attain non-vanishing amplitudes due to the quadrupole nature of
the excitation force. This fact can help to distinguish an SBGW from
other excitation mechanisms (the ``noise'').

The equation of motion for stellar oscillations in presence of
external driving by GWs is the inhomogeneous wave equation for the
velocity field $\bmath{v}$ (cf. \citealt{Siegel2011})
\begin{equation}
	\rho\left(\frac{\partial^2}{\partial t^2}-\mathcal{L}\right)\bmath{v}+\mathcal{D}(\bmath{v})=\frac{\partial}{\partial t}(\bmath{f}_{\rm{Rey}}+\bmath{f}_{\rm{entr}}+\bmath{f}_{\rm{GW}}), \label{eq:EOM}
\end{equation}
where $\rho$ is the density and $\mathcal{L}$ and $\mathcal{D}$ are
linear differential operators (see \citealt{Siegel2011} for
definitions). Furthermore, $\bmath{f}_{\rm{Rey}}$ and
$\bmath{f}_{\rm{entr}}$ denote the Reynold and entropy source terms,
respectively, and represent the driving terms due to convective
motions (see \citealt{Siegel2011} and \citealt{Samadi2001} for
definitions). In the following, we are solely interested in the driving term due
to GWs, $\bmath{f}_{\rm{GW}}$, with components given by
\begin{equation}
	f_{\rm{GW}}^i(\bmath{x},t)=\frac{1}{2}\rho x^j \frac{\partial^2
          }{\partial t^2} h^i_{\phantom{i}j}.
\end{equation}
Here, $h_{\mu\nu}$ denotes the GW tensor.\footnote{Greek indices take
  spacetime values 0,1,2,3, whereas Latin indices take spatial values
  1,2,3 only. Repeated indices are summed over.}

The velocity field
in Equation~\eqref{eq:EOM} can be expanded in terms of the complete set of
eigenfunctions $\{\bmath{\xi}_N(\bmath{x})\}$ of the operator
$\mathcal{L}$, which define the solutions
\begin{equation}
	\bmath{v}_{\rmn{hom}}(\bmath{x},t)=-\rmn{i}\omega\bmath{\xi}(\bmath{x})\rmn{e}^{-\rmn{i}\omega t}
\end{equation}
of the homogeneous problem
\begin{equation}
	\left(\frac{\partial^2}{\partial t^2}-\mathcal{L}\right)\bmath{v}_{\rm{hom}}=0,
\end{equation}
\begin{equation}
	\bmath{v}(\bmath{x},t)=\sum_N (-\rmn{i}\omega_N) A_N(t)\bmath{\xi}_N(\bmath{x})\rmn{e}^{-\rmn{i}\omega_N t}.
\end{equation}
Hence, the intrinsic complex velocity field associated with a stellar oscillation
normal mode as a function of time $t$ and position $\bm{x}$ within
the star can be written as
\begin{equation}
  \label{eq:v_N}
  \bm{v}_N(\bm{x},t)\equiv -i\omega_N
  A_N(t)\bm{\xi}_N(\bm{x})e^{-i\omega_N t},
\end{equation}
where $A_N(t)$ is a time-dependent complex amplitude,
$\bm{\xi}_N(\bm{x})$ denotes the displacement
eigenfunction, $\omega_N$ the oscillation frequency, and $N=(nlm)$ is
an abridged index denoting the eigenmode under consideration with
radial order $n$, harmonic degree $l$ and azimuthal
order $m$. In the pulsation frame of a slowly rotating star (polar
axis coincides with rotation axis) and spherical coordinates, the
displacement eigenfunction can be written as
\begin{equation}
  \bm{\xi}_N(r,\Theta,\Phi)=[\xi_{r,nl}(r) \bm{e}_r +\xi_{h,nl}(r)r\nabla]Y_{lm}(\Theta,\Phi), \label{eq:xi}
\end{equation}
which we normalize according to
\begin{equation}
  \label{eq:6}
  \int_V
  d^3x\,\rho\bm{\xi}_N^*\bm{\cdot\xi}_{N'}=I\delta_{NN'}.
\end{equation}
In the above equations, $*$ denotes complex conjugation, $V$ the
stellar volume, $I$ is a constant that we set to unit mass in cgs units,
$\xi_{r,nl}(r)$ and $\xi_{h,nl}(r)$ denote the radial and horizontal
eigenfunctions, respectively, and $Y_{lm}(\Theta,\Phi)$ is
the spherical harmonic associated with the oscillation mode.

The quantity that is directly accessible to
asteroseismic and Sun-as-a-star radial velocity measurements is the
disk-integrated, apparent surface velocity of a mode (the observed
root mean square (rms) surface velocity), which can be expressed as
(cf. also \citealt{Belkacem2009,Berthomieu1990})
\begin{equation}
  \label{eq:v_s}
  v_N=\left\langle\frac{1}{2}\lvert v_{\text{app},N}(R_0,t)\rvert^2\right\rangle^{1/2},
\end{equation}
where $\langle\rangle$ denotes
temporal average and where we defined the complex apparent surface velocity
\begin{equation}
  \label{eq:v_app}
  v_{\text{app},N}(R_0,t)=\frac{\displaystyle\int_H h(\mu)\bm{v}_N(\bm{x},t)\bm{\cdot n}\,d\Omega}{\displaystyle\int_H h(\mu)\,d\Omega}
\end{equation}
as observed in the observer's frame with spherical coordinates
$(r,\theta,\phi)$ (the direction $\theta=0$ points toward the
observer and $r=0$ corresponds to the center of the star). In
Equation~\eqref{eq:v_app}, $H$ denotes the visible half sphere
corresponding to the parameter space $\{(\theta,\phi)|
0<\theta<\pi/2,\,0<\phi<2\pi\}$, $\mu=\cos\theta$ the limb angle,
$d\Omega=d\bm{A\cdot n}=R_0^2\sin\theta \cos\theta d\theta d\phi$ the
surface element projected onto the direction of the observer, and
$R_0$ the distance between the center of the star and the layer where
the apparent surface velocity is observed. Furthermore,
$\bm{n}=\cos\theta\bm{e}_r-\sin\theta\bm{e}_\theta$ is the unit vector
at a particular position on $H$, pointing toward the observer, and
$h(\mu)$ is an appropriate limb darkening law. In other words, the
apparent surface velocity $v_N$ is the intrinsic line-of-sight
velocity at a certain layer in the atmosphere, integrated and weighted
over the visible stellar disk according to a certain limb
darkening function.

In the case of excitation by an SBGW (i.e.,
$\bmath{f}_{\rm{Rey}}=\bmath{f}_{\rm{entr}}\equiv 0$), the intrinsic
mean-square amplitude of a quadrupolar ($l=2$) stellar oscillation
mode is time independent and can be directly expressed in terms of the
normalized spectral energy density of the background as \citep{Siegel2011}:
\begin{equation}
  \label{eq:A_SBGW}
  \langle|A_N|^2\rangle=\frac{\pi^2}{25}\frac{\chi_n^2}{\eta_N\omega_N I^2}H_0^2\Omega_{\mathrm{GW}}(\omega_N),
\end{equation}
where
\begin{equation}
  \label{eq:8}
  \chi_n=\int_0^R \rho(r) r^3 [\xi_{r,n2}(r)+3\xi_{h,n2}(r)]\,dr
\end{equation}
is part of the coupling factor between the GW field
and the stellar oscillation mode. Here, $R$ denotes the stellar
radius and $\eta_N$ is the damping rate associated with mode
$N$. The normalized dimensionless function
\begin{equation}
  \Omega_{\mathrm{GW}}(\nu)\equiv\frac{1}{\rho_{\mathrm{crit}}}\frac{d\rho_{\mathrm{GW}}}{d\ln\nu} \label{eq:Om_GW}
\end{equation}
is a convenient way of characterizing the properties of an SBGW
(\citealt{Maggiore2000,Allen1999}). It measures the energy density of
GWs per unit logarithmic frequency interval in units
of the present critical energy density,
$\rho_{\mathrm{crit}}=3c^2H_0^2/8\pi G$, that is needed for a closed
geometry of the universe. Here, $H_0$ denotes the present Hubble
expansion rate, $c$ the speed of light, and $G$ the gravitational constant. 

Furthermore, it can be shown that the intrinsic amplitude $A_N$ is related to
$v_N$ as
\begin{equation}
  \label{eq:v_A}
  v_N=\frac{1}{\sqrt{2}}\langle|A_N(t)|^2\rangle^{1/2} \omega_N
  \Psi_N(R_0),
\end{equation}
where 
\begin{equation}
  \label{eq:Psi_N}
  \Psi_N(R_0)=|\alpha_{lm} \xi_{r,nl}(R_0)+\beta_{lm}
\xi_{h,nl}(R_0)|,
\end{equation}
with visibility coefficients
\begin{eqnarray}
  \alpha_{lm} &=& N_{lm}|P_{lm}(\cos\Theta_0)|u_l, \label{eq:alpha}\\
  \beta_{lm} &=& N_{lm}|P_{lm}(\cos\Theta_0)|v_l \label{eq:beta}
\end{eqnarray}
(cf. also
\citealt{Belkacem2009,Berthomieu1990,Dziembowski1977,Christensen-Dalsgaard1982}). Here,
$N_{lm}=\sqrt{(2l+1)/4\pi}\sqrt{(l-m)!/(l+m)!}$, $P_{lm}$ denote the
associated Legendre polynomials, and
\begin{eqnarray}
  u_l&=&\int_{0}^{1}  \tilde{h}(\mu)\mu^2 P_l(\mu)\,d\mu, \label{eq:u_l}\\
  v_l&=&l\int_{0}^{1} \tilde{h}(\mu)\mu[P_{l-1}(\mu)-\mu P_l(\mu)]\,d\mu, \label{eq:v_l}
\end{eqnarray}  
with $\tilde{h}(\mu)=h(\mu)/\int_0^1 h(\mu)\mu\,d\mu$ and $P_l=P_{l0}$
the Legendre polynomials.

Given an upper bound on the
apparent surface velocity $v_N$ of a quadrupolar stellar eigenmode
with frequency $\omega_N=2\pi\nu_N$ and assuming that the observed
oscillations of a star are excited at least partially by a stochastic background of
gravitational radiation as one of the driving forces,
Equations~\eqref{eq:A_SBGW} and \eqref{eq:v_A} imply an upper limit on the
normalized spectral energy density of the SBGW according to
\begin{equation}
  \label{eq:upper_bound}
  H_0^2\Omega_{\mathrm{GW}}(\nu_N)<\frac{25}{\pi^3}\mathcal{X}_N\mathcal{M}_N\frac{\eta_N}{\nu_N}v_N^2.
\end{equation}
Here, $\mathcal{X}_N\equiv 1/\chi^2_n$ is the coupling factor between
the GW field and the stellar oscillation mode,
i.e., the susceptibility of a particular mode to the
GW background. The factor $\mathcal{M}_N\equiv
I^2/\Psi_N^2(R_0)$ is the observed mode mass (the observed value for the total
interior mass of the star that is affected by the oscillation), which
takes instrumental and other observation-related
effects into account (cf. Equation~\eqref{eq:Psi_N}). We note that the left-hand side of
Equation~\eqref{eq:upper_bound} is independent of the Hubble constant and thus
independent of its uncertainty.

\section{Results for the Sun}
\label{sec:results}

The method described in the previous section is very general and
applies to any solar-like oscillator. Here, we compute an upper limit
employing our nearest star, the Sun, as a hydrodynamic detector. The
underlying solar model that we use is Model S
\citep{ChristensenDalsgaard1996}, which is extensively used as a
reference solar model (see, e.g., \citealt{Turck-Chieze2011} for a
discussion on the current status of the standard solar model). Thanks
to the very high level of agreement in the numerical results for
solar(-like) models computed with present stellar evolution codes
\citep{Lebreton2008}, any other well-fitted solar model could have
been used, such as, e.g., the CESAM model
employed by \cite{Belkacem2009}. We note that the mean quadratic
differences in the physical and seismic variables between solar
models computed with, e.g., ASTEC and CESAM are particularly small
(mean quadratic differences in the physical and thermodynamic
variables are often well below or on the order of $1\%$; differences
in the oscillation frequencies are typically less than $0.01\%$; \citealt{Lebreton2008}).

The solar oscillation modes best-suited for deducing an
upper bound on a GW background are the high-frequency
(low radial order) quadrupolar \textit{g} modes, as they are most
sensitive to such a background \citep{Siegel2011}: the corresponding
intrinsic mean-square amplitudes $\langle|A_N|^2\rangle$ are typically
orders of magnitude larger than for low radial order quadrupolar
\textit{p} modes (assuming a constant $\Omega_{\mathrm{GW}}(\nu)$),
which is mainly the result of much smaller damping rates (cf. Figure~5
in \citealt{Siegel2011}). Combined with the observed much larger
surface amplitudes, \textit{p} modes are less relevant for deducing a
tight upper bound on an SBGW and are thus not considered here (note
that the upper bound on an SBGW according to
Equation~\eqref{eq:upper_bound} scales with the squared surface velocity).

Solar \textit{g} modes have not been unambiguously detected so far,
although some \textit{g}-mode candidates have been identified in data
from the GOLF instrument aboard the \emph{SOHO} spacecraft
(\citealt{Turck-Chieze2004,Garcia2007}; for a review on quadrupolar
\textit{p} and \textit{g}-mode measurements, see, e.g., \citealt{Turck-Chieze2012}).
Their surface velocities are extremely small, because these modes are
evanescent in the convection zone. Theoretical quantitative estimates
for solar \textit{g}-mode surface velocities are highly
uncertain and differ from each other by orders of magnitude (e.g.,
\citealt{Gough1985,Bahcall1993,Kumar1996,Provost2000,Belkacem2009}),
which is predominantly due to the hypotheses made in the excitation
models concerning turbulent convection and, in particular, the choice
of the eddy-time correlation function \citep{Belkacem2009}. The
currently predicted range for quadrupolar \textit{g}-mode rms surface
velocities is $10^{-3}\,\mathrm{mm}\,\mathrm{s}^{-1}\lesssim
v_N\lesssim 1\,\mathrm{mm}\,\mathrm{s}^{-1}$ (cf. \citealt{Appourchaux2010}).

Observational upper bounds on \textit{g}-mode surface
amplitudes have been reported numerously in the literature
\citep{Appourchaux2010}. As a conservative upper bound on solar
\textit{g}-mode rms surface amplitudes we adopt
$6\,\mathrm{mm}\,\mathrm{s}^{-1}$ for any of the modes
\citep{Gabriel2002}, which is the 90\% confidence limit deduced from
Doppler measurements of the solar disk-integrated line-of-sight
velocity field as observed by the GOLF instrument. This limit is based
on Level 2 GOLF data, which were calibrated using
the method described by \cite{Ulrich2000}. Accordingly, we employ for our
calculation of visibility coefficients the limb
darkening law found by \cite{Ulrich2000},
\begin{equation}
  \label{eq:limb}
  h(\mu)=1+c_1(1-\mu)+c_2(1-\mu)^2+c_3(1-\mu)^3,
\end{equation}
where $c_1=-0.466$, $c_2=-0.06$, and $c_3=-0.29$. With this limb
darkening law, the numerical values for the coefficients $u_2$ and
$v_2$ (cf. Equations~\eqref{eq:u_l} and \eqref{eq:v_l}) are 0.324 and
0.781, respectively. We numerically
checked that the precise height in the solar atmosphere where the velocities
are observed does not significantly influence our final result. To a
very good approximation, we can therefore set $R_0=R$, where $R$ is the
stellar radius. This is to be expected for \textit{g} modes, given their small
amplitudes at the solar surface. The angle between the solar rotation
axis and the polar axis of the observer's frame as defined by the
position of the \emph{SOHO} spacecraft is $\Theta_0=83^\circ$. It is worthwhile
to note that the visibility coefficients do not depend on the
(time-dependent) azimuthal offset $\Phi_0$ between the pulsation frame
and the observer's frame. With these details, the values of the visibility
coefficients $\alpha_{lm}$ and $\beta_{lm}$ (cf. Equations~\eqref{eq:alpha}
and \eqref{eq:beta}) can be computed, which we list in Table~\ref{tab:visibility}.

A further ingredient are the damping rates. It is important to note
that radiative damping is the dominant damping mechanism for
asymptotic \textit{g} modes. Thanks to this fact, \textit{g}-mode
damping rates (excluding the first few radial orders, i.e., at least up
to $\approx 110\,\mu\textrm{Hz}$ in the case of the
Sun) can be reliably calculated with non-adiabatic oscillation
computations \citep{Dupret2002,Belkacem2009}. Above
$110\,\mu\textrm{Hz}$, time-dependent convection terms become
significant and damping rates are sensitive to the parameter $\beta$
(cf. \citealt{Grigahcene2005}) used by \cite{Belkacem2009} to model
convection-pulsation interactions. The value adopted by the latter
authors was chosen such that good agreement between theoretically
computed and observed damping rates of solar \textit{p} modes was
achieved. However, for the lack of corresponding data, this
calibration cannot be verified for solar \textit{g} modes. Although
the deviation from the radiative damping power law (see below) will
most likely not be large for the first few modes above
$110\,\mu\textrm{Hz}$, we cannot entirely trust the upper bounds
deduced above this frequency.

\begin{figure}[tb]
\centering
\includegraphics[width=0.47\textwidth]{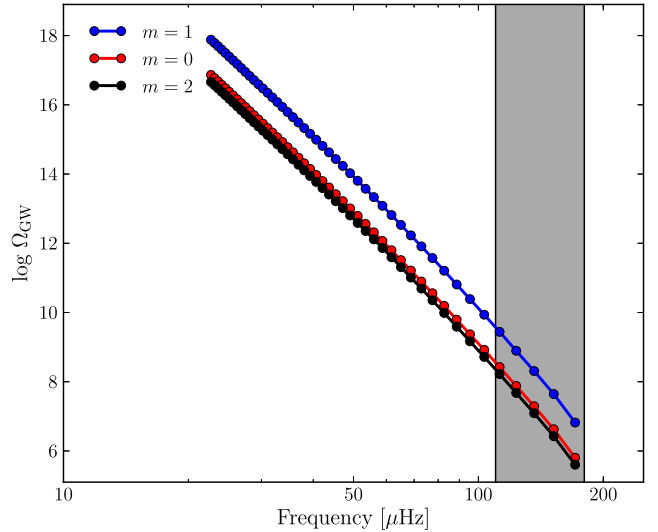}
 \caption{Upper limits on an SBGW from Sun-as-a-star data. The limits
   shown are deduced from the 90\% confidence limit on solar \textit{g}-mode
   surface amplitudes \citep{Gabriel2002} applied to asymptotic
   quadrupolar modes according to Equation~\eqref{eq:upper_bound} (dots
   indicate individual modes). Due to visibility effects, the
   degeneracy in the azimuthal order $m$ is lifted. The frequency
   regime above $110\,\mu\mathrm{Hz}$, where the damping rates cannot
   be trusted entirely, is indicated in grey (see the text for details).}
\label{fig:upper_bound}
\end{figure}

\begin{table}[b]
\caption{Values of the Visibility Coefficients
  (cf. Equations~\eqref{eq:alpha} and \eqref{eq:beta}) for Quadrupolar
  Modes According to the Limb Darkening Law Equation~\eqref{eq:limb} and an
  Inclination Angle of $\Theta_0=83^\circ$. We note that
  $\alpha_{2,-m}=\alpha_{2m}$ and $\beta_{2,-m}=\beta_{2m}$.}
\label{tab:visibility}
\centering
\begin{tabular}{ccc}
\hline\hline
$m$ & $\alpha_{2m}$ & $\beta_{2m}$ \\
\hline
0 & 0.0976 & 0.235\\
1 & 0.0303 & 0.0729\\
2 & 0.123 & 0.297\\
\hline
\end{tabular}
\end{table}

\begin{figure}
\centering
  \includegraphics[width=0.47\textwidth]{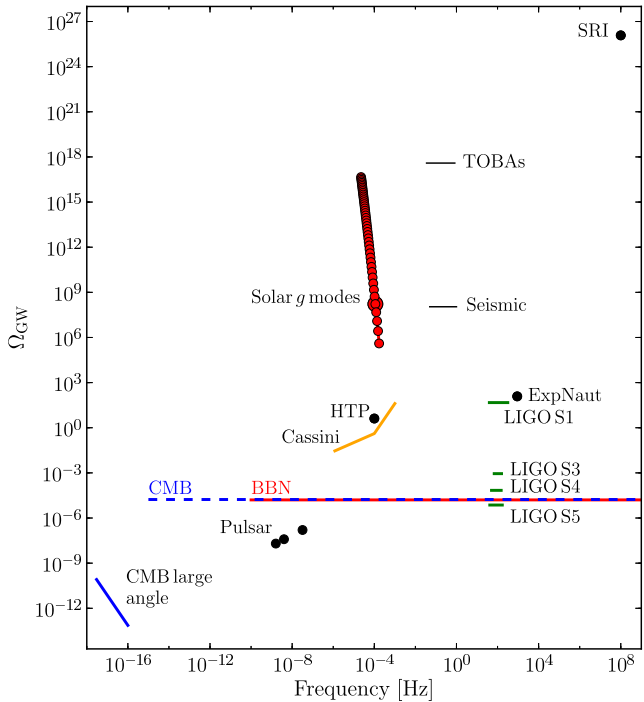}
  \caption{Comparison of present upper bounds on an SBGW and the limits
    deduced from helioseismic data of the Sun (see the text for
    details). CMB observations at large angular scales indicate
    an upper limit on the cosmological component of the SBGW at larger
    wavelengths than the horizon size at the time of decoupling
    ($3\times10^{-18}\!-\!10^{-16}\,\mathrm{Hz}$;
    \citealt{Maggiore2000}). Pulsar timing observations based on the
    residuals of the pulse arrival times yield upper bounds between
    $10^{-9}\!-\!3\times 10^{-8}\,\mathrm{Hz}$ \citep{Jenet2006}. An upper
    limit from Doppler tracking of the Cassini spacecraft is obtained
    in the frequency range $10^{-6}\!-\!10^{-3}\,\mathrm{Hz}$
    \citep{Armstrong2003}. Between $0.035\!-\!0.830\,\mathrm{Hz}$ a pair
    of TOBAs finds $\Omega_{\mathrm{GW}}< 3.9\times10^{17}$
    \citep{Shoda2014}, and a pair of synchronous recycling
    interferometers has placed $\Omega_{\mathrm{GW}}<1.2\times10^{26}$
    at $100\,\mathrm{MHz}$ \citep{Akutsu2008}. From the variance of
    orbital elements of the Hulse-Taylor binary pulsar an upper bound
    roughly one order of magnitude less stringent than the Cassini
    bound could be deduced at $10^{-4}\,\mathrm{Hz}$
    \citep{Hui2013}. Seismic data of the Earth have recently placed an
    upper bound of $\Omega_{\mathrm{GW}}<1.1\times10^{8}$ between
    $0.05\!-\!1\,\mathrm{Hz}$ \citep{Coughlin2014}. A cross-correlation
    measurement between the Explorer and Nautilus cryogenic resonant
    bar detectors yielded $\Omega_{\mathrm{GW}}<122$ at
    $907.2\,\mathrm{Hz}$ \citep{Astone1999}. Also indicated are the
    upper limits from the S1 to S5 science runs of the Earth-based
    interferometric detector LIGO around $100\,\mathrm{Hz}$, with the
    tightest bound being $\Omega_{\mathrm{GW}}(\nu)<7.3\times 10^{-6}$
    between $41.5\!-\!169.25\,\mathrm{Hz}$ at 95\% confidence
    \citep{Abbott2009}. Indirect bounds can be deduced from BBN and
    CMB data \citep{Maggiore2000,Cyburt2005,Smith2006}, which
    constrain the integrated total energy density of the cosmological
    component of the SBGW in the indicated frequency ranges (see the
    text for details).}
\label{fig:upper_bounds}
\end{figure}

Figure \ref{fig:upper_bound} shows the upper bounds on an SBGW as
deduced from quadrupolar ($l=2$) asymptotic
solar \textit{g} modes according to Equation~\eqref{eq:upper_bound}, in which we
have assumed $H_0=70\,\mathrm{km}\,\mathrm{s}^{-1}\,\mathrm{Mpc}^{-1}$ \citep{Komatsu2011}. As
can be seen from Figure~\ref{fig:upper_bound}, the upper bounds are
tightest at high frequencies and they
follow a fairly regular power law, which can be understood from an
asymptotic analysis of the quantities appearing in
Equation~\eqref{eq:upper_bound}. The quantities $\chi_n^2$ and $\eta_N$ are
known to show power-law behavior as a function of frequency for
asymptotic \textit{g} modes \citep{Siegel2011,Belkacem2009}. Here, we
also find a power law for $\Psi_N^2(R_0)$ in the case of asymptotic \textit{g}
modes. Therefore, according to Equation~\eqref{eq:upper_bound}, the upper
bound on $\Omega_{\mathrm{GW}}(\nu)$ is also of power-law
form. We note that the degeneracy in the azimuthal
order $m$ of the modes is lifted due to visibility effects, which are
encoded in the quantity $\Psi_N^2(R_0)$ through the visibility
coefficients $\alpha_{lm}$ and $\beta_{lm}$. The regime above $\approx
110\,\mu\textrm{Hz}$, in which we
cannot entirely trust the damping rates (see above), is indicated in
gray. The tightest bound that can still be reliably deduced is
$\Omega_{\mathrm{GW}}< 1.7\times10^8$ at $0.112\,\mathrm{mHz}$, while the
tightest overall bound is $\Omega_{\mathrm{GW}}< 4.0\times10^5$ at
$0.171\,\mathrm{mHz}$. These values are reduced by a factor of four if an
upper bound of $3\,\mathrm{mm}\,\mathrm{s}^{-1}$ on the apparent
\textit{g}-mode surface velocities according to \citet{Turck-Chieze2004} is
used.\footnote{\citet{Turck-Chieze2004} estimated the amplitudes
    of the \textit{g}-mode candidates they reported to $2\pm
    0.9\,\mathrm{mm}\,\mathrm{s}^{-1}$.}
As evident from Figure~\ref{fig:upper_bound}, the tightest bounds are
obtained for \textit{g} modes with $l=m=2$, which
are compared to other observational constraints on an SBGW in
Figure~\ref{fig:upper_bounds}. In the latter figure, we mark
the former of the aforementioned limits with a large dot.

\section{Discussion and Conclusions}
\label{sec:discussion}

The frequency range $10^{-4}\!-\!10$\,Hz, which
is particularly interesting from the astrophysical point of view, has essentially
remained unexplored until today in terms of strong bounds. However, several low-frequency
antennas are currently being proposed, such as NGO (a.k.a. eLISA),
DECIGO, or TOBA, with final sensitivities that are claimed to reach
$\Omega_{\mathrm{GW}}\sim 10^{-9},\, 10^{-15}$ and
$\Omega_{\mathrm{GW}}\sim 10^{-8}$, respectively
\citep{Amaro-Seoane2013,Kawamura2011,Ishidoshiro2011}. In this
frequency range, cosmological backgrounds (e.g., from cosmic strings)
could be outshined by the astrophysical backgrounds from binary
neutron stars and galactic as well as extragalactic white dwarfs
binaries \citep{Regimbau2011,Schneider2010,Amaro-Seoane2013}. At lower
frequencies where our method can still be applied, the astrophysical
background due to supermassive black hole binaries becomes important
(e.g., \citealt{Sesana2004}). The method presented in this paper makes
the aforementioned frequency range accessible by providing a
possibility to place direct bounds on an SBGW at $\mu$Hz and mHz
frequencies with asteroseismic data. Despite some partial overlap
with the Cassini band, this method can somewhat bridge the gap between
the Cassini range and the bounds from Earth-based interferometers,
possibly improving on the Cassini limits (see below). It is important
to point out that there are indirect upper limits in this frequency
range, which are deduced from CMB and BBN data due to the fact that a
larger amplitude of the SBGW would have altered the observed
abundances of the light nuclei created during BBN; analogously, a
larger amplitude would have also modified the CMB and matter power
spectra. However, these are indirect bounds in the sense that they
constrain the integrated total energy density
\begin{equation}
  \Omega_\mathrm{GW}=\int\Omega_{\mathrm{GW}}(\nu) d\ln\nu  \label{eq:indirect_bounds}
\end{equation}
over the frequency ranges indicated in Figure~\ref{fig:upper_bounds} as
$\Omega_{\mathrm{GW,BBN}}<1.6\times 10^{-5}$ and
$\Omega_{\mathrm{GW,CMB}}<1.7\times 10^{-5}$, respectively
\citep{Maggiore2000,Cyburt2005,Smith2006}. Furthermore, they solely
constrain primordial GWs, which already existed at the
time when, respectively, the CMB was formed
and the nucleosynthesis took place. In particular, these bounds cannot
constrain the astrophysically generated GWs, which were produced at later times and which are still
generated today. In contrast to these indirect limits, our method
reported here yields direct upper bounds that additionally
constrain the astrophysical component of the stochastic
GW background. They complement the direct upper
limits in other frequency bands and the integrated, indirect upper
limits in the same frequency regime. Furthermore, unlike most of the other
direct bounds, our method does not assume
a certain spectral shape for the function $\Omega_{\mathrm{GW}}(\nu)$,
i.e., it applies to an arbitrary function $\Omega_{\mathrm{GW}}(\nu)$,
regardless of the global shape of $\Omega_{\mathrm{GW}}(\nu)$. 

The upper bound given
by Equation~\eqref{eq:upper_bound} is highly sensitive to local
(i.e., seismic; cf. the wide range in sensitivity in
Figures~\ref{fig:upper_bound} and \ref{fig:upper_bounds}) and global stellar
properties and to the observed surface velocities of the modes. Space
missions like \emph{CoRoT} and \textit{Kepler} recorded asteroseismic
intensity data for a wide range of stellar mass, radius and effective
temperature. Furthermore, radial velocity measurements of stellar
oscillations from ground exist. The precision of helioseismic
measurements might not be achieved by asteroseismology. In particular,
from solar observations it can be concluded that the
signal-to-noise ratio of intensity variations are expected to be an
order of magnitude lower than for velocity measurements
\citep{Nigam1998}, and radial velocity measurements might not yet
achieve a precision on the order of mm/s. However, these limitations
could be overcompensated by several orders of magnitude due to the
dependence of Equation~\eqref{eq:upper_bound} on, e.g., stellar mass and
radius~\citep{Siegel2011}, and stellar modelling can guide
observations to find optimal targets. Still, a conversion from intensity to
velocity amplitudes is required before using intensity data in
Equation~\eqref{eq:upper_bound}, but this might be, e.g., obtained
empirically from comparisons of intensity and velocity measurements,
as, for instance, carried out by \emph{SOHO} for the Sun \citep{Nigam1998} or
between \textit{Kepler} intensity and follow-up radial velocity
data for the stars. Therefore, the existing and up-coming space-based
missions, as well as ground-based facilities like SONG
\citep{Grundahl2009}, offer a unique possibility to focus on targets
with optimized sets of stellar parameters in terms of detector
sensitivity and to possibly employ these stars as a large array of low-frequency
antennas for the SBGW at $\mu\mathrm{Hz}$ and $\mathrm{mHz}$
frequencies by using the method presented here.

\vspace{0.5cm}
The authors thank K. Belkacem and R. Samadi for valuable discussions
and K. Belkacem for sharing the solar \textit{g}-mode damping rates published
in \cite{Belkacem2009}.


\end{document}